\RequirePackage{fix-cm}
\documentclass[twocolumn]{svjour3}          
\smartqed  
\usepackage{graphicx}
\usepackage{latexsym}
\usepackage{amsmath}
\usepackage{amssymb}
\usepackage{multirow}
\usepackage[utf8]{inputenc}
\journalname{Brazilian Journal of Physics}

\begin{document}

\title{Random Sampling of Quantum States: A Survey of Methods}
\subtitle{And Some Issues Regarding the Overparametrized Method}

\author{Jonas Maziero }

\institute{Jonas Maziero \at Departamento de F\'isica, Centro de Ci\^encias Naturais e Exatas, Universidade Federal de Santa Maria, Avenida Roraima 1000, 97105-900, Santa Maria, RS, Brazil \at Instituto de F\'isica, Facultad de Ingenier\'ia, Universidad de la Rep\'ublica, J. Herrera y Reissig 565, 11300, Montevideo, Uruguay \\ \email{jonas.maziero@ufsm.br}}

\date{Received: date / Accepted: date}

\maketitle

\begin{abstract}
The numerical generation of random quantum states (RQS) is an important procedure for investigations in quantum information science. Here we review some methods that may be used for performing that task. We start by presenting a simple procedure for generating random state vectors, for which the main tool is the random sampling of unbiased discrete probability distributions (DPD). Afterwards the creation of random density matrices is addressed. In this context we first present the standard method, which consists in using the spectral decomposition of a quantum state for getting RQS from random DPDs and random unitary matrices. In the sequence the Bloch vector parametrization method is described. This approach, despite being useful in several instances, is not in general convenient for RQS generation. In the last part of the article we regard the overparametrized method (OPM) and the related Ginibre and Bures techniques. The OPM can be used to create random positive semidefinite matrices with unit trace from randomly produced general complex matrices in a simple way that is friendly for numerical implementations. We consider a physically relevant issue related to the possible domains that may be used for the real and imaginary parts of the elements of such general complex matrices. Subsequently a too fast concentration of measure in the quantum state space that appears in this parametrization is noticed.

\end{abstract}

\maketitle
\section{Introduction}

About three decades ago Paul Benioff \cite{Benioff_jsp,Benioff_prl}
and Richard Feynman \cite{Feynman_SPC,Feynman_QMC} envisaged a computer
whose basic constituents could be in a complex quantum superposition
state. In the last few years we have been witnessing astonishing theoretical
and experimental developments in quantum computing and quantum simulation
\cite{Bennett_QIC,Laflamme_QC,Nori_QS}, and also in others sub-areas
of quantum information science \cite{Renner_QKD,Nori_QB,Jarzynski_QT,Petruccione_QML},
with experimental implementations already going beyond the best present
classical capabilities \cite{Eisert_Spup}. These are the first sights
of what will turn out to be a revolution in our science and technology
\cite{Preskill_Fut,Aaronson_Fut}. 

Nevertheless, before that can in fact become a reality, we still
have much to understand concerning quantum systems with many degrees
of freedom. One important tool for accomplishing this task is the
generation and analysis of RQS \cite{James_rand_r-1,Ramos_rand_r,Adesso_Adisc,Plastino,Genovese_rqs,Rau_rqs,Wang_rqs,Boyd_,Boyd,Wootters,Hall,Nechita,Nadal,Hamma,Agarwal,Cunden,Hastings},
which will have an analogous role to that that random numbers
have in classical stochastic theories \cite{Jaynes_Prob,Binder_MC,Cover-Thomas,Devore_Prob_App}.
The parametrization of quantum states \cite{James_rand_r-1,Petruccione_PofU-1,Fritzche4}
is the initial step towards generating them numerically and is one of the
main topics of this survey, which is organized in the following manner. In Sec. \ref{pure} we consider the generation of random pure states, for which the availability of unbiased random discrete probability distributions is indispensable and is hence also recapitulated. The remainder of the article is dedicated to the creation of general random density matrices. In Sec. \ref{standard} the standard method is described together with the Hurwitz parametrization for unitary matrices, which used in its implementation. Subsequently, in Sec. \ref{Bloch}, the Bloch vector parametrization, though impractical for RQS generation, is regarded for completeness. The last part of the article, the Sec. \ref{OPM}, is dedicated to present and investigate some issues regarding the overparametrized and related methods. In Sec. \ref{issue} we discuss unwanted physical consequences of the ranges usually used in the literature for the real and imaginary parts of the elements of the general complex matrices involved in this method and present a simple solution for the problem. In Sec. \ref{concentration} we report an important possible drawback of the
OPM regarding its use for random sampling in the quantum state space: its too fast concentration of measure. We discuss the Ginibre and Bures methods in Sec. \ref{bures}. A brief summary of the article is presented in Sec. \ref{conclusions}.

\section{Pure States}
\label{pure}

When there is no classical uncertainty about the state of a quantum system, it is represented by a vector in a Hilbert space $\mathcal{H}$. For discrete systems, $\mathcal{H}$ is simply $\mathbb{C}_{d}$ with the inner product between any two of its vectors defined as $\langle\psi | \phi\rangle:=| \psi \rangle^{\dagger} |\phi\rangle=\sum_{j=1}^{d}\psi_{j}^{*}\phi_{j}$, where $z^{*}$ is the complex conjugate of $z$ and $d$ is the system dimension. Here we use the standard notation of Dirac for vectors and, for a generic
matrix $A$, we denote $A^{\dagger}$ as its adjoint (conjugate transpose). Any state $|\psi\rangle\in\mathcal{H}$ can be written as a linear combination of the vectors of any basis. One base of special interest is the computational or standard basis: $|c_{1}\rangle=[1 \mbox{ } 0 \mbox{ } \cdots \mbox{ } 0]^{T}$, $|c_{2}\rangle=[0 \mbox{ } 1 \mbox{ } \cdots \mbox{ } 0]^{T}$, $\cdots$, $|c_{d}\rangle=[0 \mbox{ } 0 \mbox{ } \cdots \mbox{ } 1]^{T}$, in terms of which
\begin{equation}
 |\psi\rangle =\textstyle{\sum_{j=1}^{d}}\psi_{j}|c_{j}\rangle,
\end{equation}
with $\psi_{j} = \langle c_{j}|\psi\rangle$. Above, $X^{T}$ denotes the transpose of the matrix $X$.

The Born's probabilistic interpretation of the state vector $|\psi\rangle$ requires its normalization:
\begin{equation}
||\vec{\psi}||:=\sqrt{\langle\psi | \psi\rangle}=\sqrt{\textstyle{\sum_{j=1}^{d}}|\psi_{j}|^{2}}=1.
\end{equation}
Thus, as the numbers $|\psi_{j}|^{2}$ are non-negative and sum up to one, they form a probability distribution $p_{j}:=|\psi_{j}|^{2}$. Using $\psi_{j}=|\psi_{j}|\exp(i\theta_{j})$ one can write
\begin{equation}
 |\psi\rangle =  \textstyle{\sum_{j=1}^{d}} \sqrt{p_{j}}\exp(i\theta_{j})|c_{j}\rangle, 
 \label{eq_psi}
\end{equation}
with the phases $\theta_{j}\in[0,2\pi]$.

Now we recall that if we have access to a random number generator yielding random numbers with uniform distribution in $[0,1]$, an unbiased random discrete probability distribution (RDPD) \cite{Fritzche4,Vedral_Prob_trigo} can be generated as follows \cite{Maziero_pRPV}. First we create a biased RDPD generating $q_{1}$ in the interval $[0,1]$ and $q_{j}$ in $[0,1-\sum_{k=1}^{j-1}p_{k}]$ for $j=2,\cdots,d$. Then we use a random permutation of $\{1,\cdots,d\}$, let us call it $\{k_{1},\cdots,k_{d}\}$, and define the unbiased RDPD as
\begin{equation}
\{p_{1},\cdots,p_{d}\}:=\{q_{k_{1}},\cdots,q_{k_{d}}\}.
\end{equation}

The unbiased RDPD generated in this way and $d$ independent random phases $\theta_{j}$ are then applied to generate a random pure state. It is worth observing that there will be no privileged direction in $\mathcal{H}$ only because the RDPD is unbiased. This pure state generation procedure gives $|\psi\rangle$ distributed with a Haar measure. Another manner of obtaining samples with similar properties is by using the rows or columns of random unitary matrices, which we shall discuss in the next section.

\section{Standard method}
\label{standard}

The states of a $d-$level quantum system are described, in the most
general scenario, by a density matrix $\rho$ \cite{Nielsen_Book,Wilde_Book},
which is a Hermitian positive semidefinite matrix (notation: $\rho\ge0$) with
unit trace ($\mathrm{Tr}(\rho)=1$). Any such matrix can be written
in the form of a spectral decomposition:
\begin{equation}
\rho={\textstyle \sum_{j=1}^{d}}r_{j}|r_{j}\rangle\langle r_{j}|,\label{eq:dm1}
\end{equation}
with the real eigenvalues of $\rho$ being nonnegative ($r_{j}\ge0$
for all $j=1,\cdots,d$) and summing up to one ($\sum_{j=1}^{d}r_{j}=1$).
That is to say, $\{r_{j}\}_{j=1}^{d}$ is a probability distribution
\cite{Jaynes_Prob,Devore_Prob_App}. The eigenvectors of $\rho$,
$\{|r_{j}\rangle\}_{j=1}^{d}$, form an orthonormal basis for the
vector space $\mathbb{C}^{d}$, i.e., $\langle r_{j}|r_{k}\rangle =\delta_{jk}$
and $\sum_{j=1}^{d}|r_{j}\rangle\langle r_{j}|=\mathbb{I}_{d}$, where
$\mathbb{I}_{d}$ is the $d\mathrm{x}d$ identity matrix.

Let us briefly look over the number of real parameters needed for
a complete description of an arbitrary density matrix. In order to
describe the probability distribution $\{r_{j}\}_{j=1}^{d}$, the
eigenvalues of $\rho$, we need $d-1$ real numbers. Besides, as any
two bases for the vector space $\mathbb{C}^{d}$ are connected by
an unitary matrix $U$ (i.e., $UU^{\dagger}=\mathbb{I}_{d}$), one
can write
\begin{equation}
|r_{j}\rangle=U|c_{j}\rangle,
\end{equation}
for $j=1,\cdots,d$, with $\{|c_{j}\rangle\}_{j=1}^{d}$ being the
computational basis, as shown in Sec. \ref{pure}. Therefore the bases $\{|r_{j}\rangle\}_{j=1}^{d}$ is completely determined by $U$. Once $d^{2}-d$ real parameters
are sufficient to specify completely an
arbitrary unitary matrix $U$ with dimensions $d\mathrm{x}d$ \cite{Petruccione_PofU-1}, it
follows that $d^{2}-1$ independent real parameters are sufficient
for a thorough description of any density matrix.

From the last two paragraphs, we see that the numerical generation
of a RQS (using the density matrix as written in Eq. (\ref{eq:dm1}))
can be cast in terms of the creation of a RDPD and of a random
unitary matrix (RU) \cite{Petruccione_PofU-1,Stewart_rand_U,Zyczkowski_U-1,Lloyd_rand_U}. We call it the \emph{standard method} because it would be a natural first choice giving the defining properties of a density matrix. Moreover, it utilizes as few real parameters as possible. This is a nice characteristic in the view that for doing some statistics with RQS, one in general needs to generate many of them, what can be a very time-consuming task for large values of the system dimension $d$.

From the several possibilities available \cite{Petruccione_PofU-1},
in this article we choose the Hurwitz parametrization for generating
RUs \cite{Zyczkowski_U-1}. In this parametrization one writes any $d\mathrm{x}d$ unitary matrix $U$ in terms of unitaries  
\begin{equation}
U^{(i,j)}(\phi_{ij},\psi_{ij},\chi_{ij})
\end{equation}
in bi-dimensional sub-spaces. The non-null elements of such elementary transformations are:
\begin{eqnarray}
 &  & U_{k,k}^{(i,j)}=1\mbox{ for }k=1,\cdots,d\mbox{ and }k\ne i,j\mbox{; } \nonumber \\
 &  & U_{i,i}^{(i,j)}=\cos(\phi_{ij})\exp(i\psi_{ij})\mbox{; }U_{i,j}^{(i,j)}=\sin(\phi_{ij})\exp(i\chi_{ij})\mbox{;} \nonumber \\
 &  & U_{j,i}^{(i,j)}=-(U_{i,j}^{(i,j)})^{*}\mbox{; }U_{j,j}^{(i,j)}=\cos(\phi_{ij})\exp(-i\psi_{ij}).
\end{eqnarray}
 
A general unitary transformation, for a $d$-level quantum system, can then be written as
\begin{equation}
U=\exp(i\alpha)U_{1}U_{2}U_{3}\cdots U_{d-1},
\end{equation}
with the sub-matrices being
\begin{eqnarray}
 &  & U_{1}=U^{(1,2)}(\phi_{12},\psi_{12},\chi_{12}), \nonumber \\
 &  & U_{2}=U^{(2,3)}(\phi_{23},\psi_{23},0)U^{(1,3)}(\phi_{13},\psi_{13},\chi_{13}), \nonumber \\
 &  & U_{3}=U^{(3,4)}(\phi_{34},\psi_{34},0)U^{(2,4)}(\phi_{24},\psi_{24},0) \nonumber  \\ && \hspace{1.0cm} U^{(1,4)}(\phi_{14},\psi_{14},\chi_{14}), \nonumber \\
 &  & \vdots \nonumber \\
 &  & U_{d-1}=U^{(d-1,d)}(\phi_{d-1,d},\psi_{d-1,d},0) \nonumber \\ && \hspace{1.2cm} U^{(d-2,d)}(\phi_{d-2,d},\psi_{d-2,d},0)\cdots \nonumber \\ && \hspace{1.2cm} U^{(1,d)}(\phi_{1,d},\psi_{1,d},\chi_{1,d}).
\end{eqnarray}
The random numbers appearing in the last equations are distributed uniformly in the following ranges of values:
\begin{eqnarray}
 &  & 0\le\alpha<2\pi\mbox{; }0\le\psi_{ij}<2\pi\mbox{; }0\le\chi_{ij}<2\pi\mbox{;}\\
 &  & \phi_{ij}=\arcsin(\xi_{ij}^{1/2i})\mbox{, }0\le\xi_{ij}<1\mbox{, }i=1,2,\cdots,d-1 \nonumber .
\end{eqnarray}

 It is worthwhile mentioning that, although not advantageous, it is possible to use the rows or columns of such a random unitary matrix as random state vector.
 
\section{Bloch vector parametrization method}
\label{Bloch}

The Hermitian-traceless-orthonormal generators of the special unitary group $SU(d)$, $\Gamma_{j}$ ($j=1,\cdots,d^{2}-1$), and $\mathbb{I}_{d}$ can be used as a basis in terms of which we can write any $d\mathrm{x}d$ density matrix in the so called Bloch vector parametrization \cite{Petruccione_PofU-1}:
\begin{equation}
\rho = \sum_{j=0}^{d^{2}-1} \gamma_{j}\Gamma_{j}, 
\label{eq:Bloch}
\end{equation}
where $\vec{\gamma}=(\gamma_{1},\cdots,\gamma_{d^{2}-1})$ is the so called Bloch's vector. One can use $\mathrm{Tr}(\rho)=1$ to see that $\gamma_{0}=1/d$ and $\mathrm{Tr}(\Gamma_{j}\Gamma_{k})=2\delta_{jk}$ to show that the coefficients in Eq. (\ref{eq:Bloch}) are half of the mean values of the aforementioned generators of $SU(d)$, i.e., $\gamma_{j}=2^{-1}\langle \Gamma_{j} \rangle\in\mathbb{R}$. 

For producing random quantum states using the Bloch vector parametrization, $d^{2}-1$ real random numbers $\gamma_{j}$ must be generated. The main difficult here is that for $d\ge 3$ there is no known explicit determination of the range of values for the parameters $\gamma_{j}$ that will lead to a physical state. Thus, given a basis for $SU(d)$, we may use the spectrum of each $\Gamma_{j}$ to determine the range from which we shall sample the corresponding $\gamma_{j}$. In the context of RQS generation, one attractive choice for the generators of $SU(d)$ are the generalized Gell Mann matrices:
\begin{eqnarray}
&& |c_{j}\rangle\langle c_{k}| + |c_{k}\rangle\langle c_{j}|   \mbox{ for } 1\leq j < k \leq d, \label{Gen_SUd_1} \\
&& -i | c_{j} \rangle\langle c_{k}| +i | c_{k} \rangle\langle c_{j} |  \mbox{ for }   1\leq j < k \leq d, \label{Gen_SUd_2} \\
&& \frac{ \textstyle \sum_{j=1}^{l}|c_{j}\rangle\langle c_{j}| -l | c_{l+1}\rangle\langle c_{l+1} | }{\sqrt{l(l+1)/2}}  \mbox{ for } 1\leq l \leq d-1.
\label{Gen_SUd_3}
\end{eqnarray}
A simple analysis shows that for the generators in Eqs. (\ref{Gen_SUd_1}) and (\ref{Gen_SUd_2}) we have $\gamma_{j}\in[-1/2,1/2]$ while for those in Eq. (\ref{Gen_SUd_3}) $\gamma_{j}\in[-\sqrt{l/(2(l+1))},1/\sqrt{2l(l+1)}]$.

Although the condition $\mathrm{Tr}(\rho)=1$ is promptly satisfied, after generating the entire Bloch's vector we must yet do a positivity test. This task requires much computational time, what makes this method impractical for the task under scrutiny here.  

\section{Overparametrized method}
\label{OPM}

The basic motivational idea for this method comes from the simple
observation that, for any complex matrix $A=(A_{jk})$, we have: $\langle\psi|A^{\dagger}A|\psi\rangle=||A|\psi\rangle||^{2}=|||\phi\rangle||^{2}\ge0$,
where $|\psi\rangle$ is any vector of $\mathbb{C}^{d}$ and $|||\phi\rangle|| := \sqrt{\langle\phi|\phi\rangle}$
is the Euclidean norm of the vector $|\phi\rangle\in\mathbb{C}^{d}$.
That is to say, for a general complex matrix $A$, the matrix $A^{\dagger}A$
is guaranteed to be positive semidefinite ($A^{\dagger}A\ge0$). Thus,
if $A$ is normalized, i.e., if we define
\begin{equation}
\mathcal{A} := \frac{A}{||A||_{2}},
\end{equation}
it is possible to write a valid density operator as:
\begin{equation}
\rho=\mathcal{A}^{\dagger}\mathcal{A}.\label{eq:dm2}
\end{equation}
Above $||A||_{2} := \sqrt{\langle A|A\rangle}$ is the \emph{Hilbert-Schmidt
norm} of $A$, with $\langle A|B\rangle := \mathrm{Tr}(A^{\dagger}B)$
being the Hilbert-Schmidt inner product between the matrices $A$
and $B$ \cite{Wilde_Book}. 

The simple formula for $\rho$ in Eq. (\ref{eq:dm2}) has found applications
in quantum information science \cite{James_rand_r-1,Boyd,Fritzche4,Englert1,Englert2}.
Once the complete description of a general complex $d\mathrm{x}d$
matrix $A$ requires $2d^{2}$ real parameters, one notes that this
parametrization, despite being simple and friendly for numerical implementations,
uses more real numbers than necessary, as discussed above. Thus it
is dubbed as the \emph{overparametrized method}. The numerical generation
of RQS via this method is further explained in the next sub-section.

\subsection{An issue on the domains of $\mathrm{Re}(A_{jk})$ and $\mathrm{Im}(A_{jk})$}

\label{issue}

Let us start our analysis of the production of RQS via the overparametrized
method by considering the simplest quantum system, a two-level system
also known as quantum bit, or qubit for short. The advantage of using
this system as our starting point is that it can be visualized straightforwardly
in the $\mathbb{R}^{3}$. For that purpose we simply need to write
a density operator $\rho$ using the $2\mathrm{x}2$ identity matrix
$\mathbb{I}_{2}$ and the Pauli matrices $\sigma_{j}$ ($j=1,2,3$)
as a basis (the case $d=2$ in the Bloch method):
\begin{equation}
\rho=2^{-1}\mathbb{I}_{2}+{\textstyle \sum_{j=1}^{3}}2^{-1}x_{j}\sigma_{j},\label{eq:rho_BB}
\end{equation}
where $x_{j}=\mathrm{Tr}(\rho\sigma_{j})$ is the value of the component
of the system's ``polarization'' in the direction $j=1,2,3\equiv x,y,z$.
The real numbers $(x_{1},x_{2},x_{3})\equiv(x,y,z)$ are used as the
Cartesian coordinates in $\mathbb{R}^{3}$. Enforcing the $\rho$
in Eq. (\ref{eq:rho_BB}) to be a density matrix leads to the following
restrictions \cite{Nielsen_Book}: $-1\le x_{j}\le1$ and $\sum_{j=1}^{3}x_{j}^{2}\le1$.
Therefore the points $(x_{1},x_{2},x_{3})$ must lie within a ball
with radius equal to one and centered at $(0,0,0)$, known as the
\emph{Bloch's ball} (BB).

There are several functions one may be interested in when working
in quantum information science. Some relevant examples are quantifiers
for total correlation \cite{Maziero_dist_MI}, quantum entanglement \cite{Davidovich_RevE}, quantum discord \cite{Lucas_RevD}, quantum coherence \cite{Plenio_QC,Girolami_QC},
and quantum channel capacities \cite{Giovannetti_Channels}. All of
these quantities can, in general, be defined using distance measures
in the quantum state space.
For our purposes in this article, the Hilbert-Schmidt distance
(HSD) fits well. The HSD between two density matrices $\rho$ and
$\zeta$ is defined as the Hilbert-Schmidt norm of their subtraction
\cite{Nielsen_Book,Wilde_Book}:
\begin{equation}
d_{hs}(\rho,\zeta) := ||\rho-\zeta||_{2}.
\end{equation}
If $\lambda_{j}$ are the real eigenvalues of the Hermitian matrix
$\rho-\zeta$, then 
\begin{equation}
d_{hs}(\rho,\zeta)=\sqrt{\mathrm{Tr}\left((\rho-\zeta)^{\dagger}(\rho-\zeta)\right)}=\sqrt{{\textstyle \sum_{j=1}^{d}}\lambda_{j}^{2}}.
\end{equation}

\begin{figure}
\begin{centering}
\includegraphics[scale=0.35]{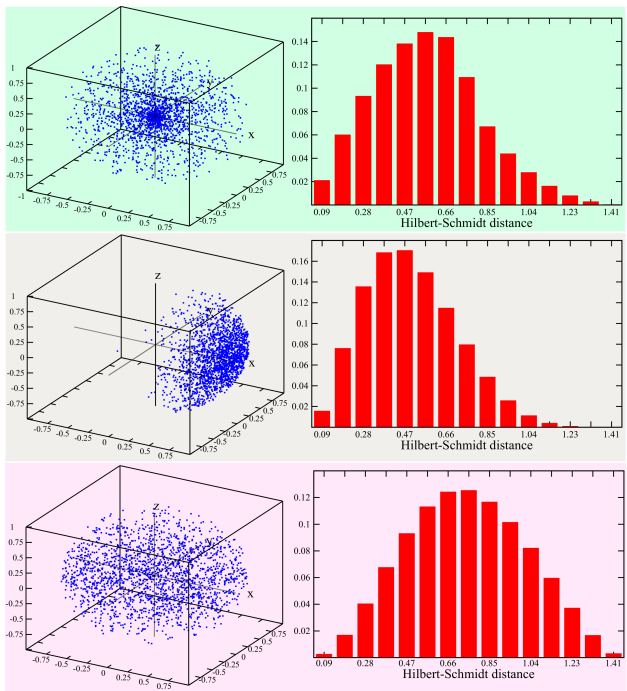}
\par\end{centering}
\protect\caption{(color online) One the left is presented the distribution in the Bloch's
ball of two thousand pseudo-random one-qubit states generated using
standard method (upper green panel) and using the overparametrized
method with the ranges for the matrices elements as utilized in Refs.
\cite{James_rand_r-1,Fritzche4}, i.e., $A_{jk}^{r}\mbox{, }A_{jk}^{i}\in[0,1]$
(gray panel in the middle) and with $A_{jk}^{r}\mbox{, }A_{jk}^{i}\in[-1,1]$
(pink panel at the bottom). One the right hand side is shown the probability
distribution for the Hilbert-Schmidt distance of one million pseudo-random quantum states
generated using the corresponding method (see the text for more details).}
\label{fig1}
\end{figure}

For the calculations involved in this article, the so called Mersenne Twister
method \cite{Matsumoto_MT} is used as the pseudo-random number generator
(pRNG) and the LAPACK subroutines \cite{LAPACK} are utilized for
computing eigenvalues. With these tools at hand, when the standard
method described in Sec. \ref{standard} is applied for generating
one-qubit pseudo-random quantum states, the distribution of such states in the Bloch's
ball and the histogram for the probability of the possible values
of HSD are those shown in the upper green panel of Fig. \ref{fig1}.
It is worth mentioning that the higher density of states observed
closer to the center of this figure can be understood by noticing
that the direction in $\mathbb{R}^{3}$ defined by $U$ is random
and that $r_{1}$ and $r_{2}=1-r_{1}$ are uniformly distributed in
the interval $[0,1]$.

Let us consider the same kind of computation, but applying now the
overparametrized method for generating the pseudo-RQS. For that purpose
the pRNG can be utilized for the sake of obtaining pseudo-random numbers
for generating the real, 
\begin{equation}
\mathrm{Re}(A_{jk}) =: A_{jk}^{r},
\end{equation}
 and imaginary, 
\begin{equation}
\mathrm{Im}(A_{jk}) =: A_{jk}^{i},
\end{equation}
 parts of the matrix elements of $A=(A_{jk})$. The first issue we
want to deal with here is with regard to the domains that one may
use for those numbers. For instance, we can follow Refs. \cite{James_rand_r-1,Fritzche4,Miszczak} and generate the matrix elements $A_{jk}$ using
uniformly distributed random numbers and setting
\begin{equation}
A_{jk}^{r}\mbox{, }A_{jk}^{i}\in[0,1].
\end{equation}
As shown at the right hand side of the gray panel at the middle of
Fig. \ref{fig1}, the probability distribution for the HSD obtained
in this way is, to some extent, qualitatively similar to that obtained
using the standard method. This may lead to the impression that our
choice for the domain of the matrix elements is fine. However, a rapid
inspection of the distribution of states in the Bloch's ball obtained
using the OPM with $A_{jk}^{r}\mbox{, }A_{jk}^{i}\in[0,1]$ reveals
a misfortune. Even though the polarization in the $y$ and $z$ directions
have approximately equal chance to be positive or negative, only positive
values for the polarization in the $x$ direction are generated. There
is no need to say that such a restriction over the possible values
of physical observables of the system is not a desirable feature for
a method supposed to generate random quantum states.

We notice that a simple solution for this problem is generating the
matrix elements $A_{jk}$ with
\begin{equation}
A_{jk}^{r}\mbox{, }A_{jk}^{i}\in[-1,1].
\end{equation}
With this change, for this case, the distribution of states in the
BB becomes even more uniform than that that we get using the standard
method, as shown in the pink panel at the bottom of Fig. \ref{fig1}.
We want to emphasize already at this point that increasing the range
of values for $A_{jk}^{r}$ and $A_{jk}^{i}$ does not causes any
significant modification neither of these results nor of those that
shall be reported in the next sub-section.

\subsection{A too fast concentration of measure for the OPM}

\label{concentration}

In the previous sub-section we showed that the application of the overparametrized
method with the real and imaginary parts of $A_{jk}$ drawn randomly
and uniformly from the interval $[-1,1]$ yields an uniform distribution
of one-qubit pseudo-random density matrices. This encouraging result
leads naturally to the question of if such a scheme can be applied
appropriately for random sampling in high-dimensional quantum systems.
In this section we investigate this question and present strong evidences
for answering it in the negative. 

\begin{figure}[b]
\begin{centering}
\includegraphics[scale=0.51]{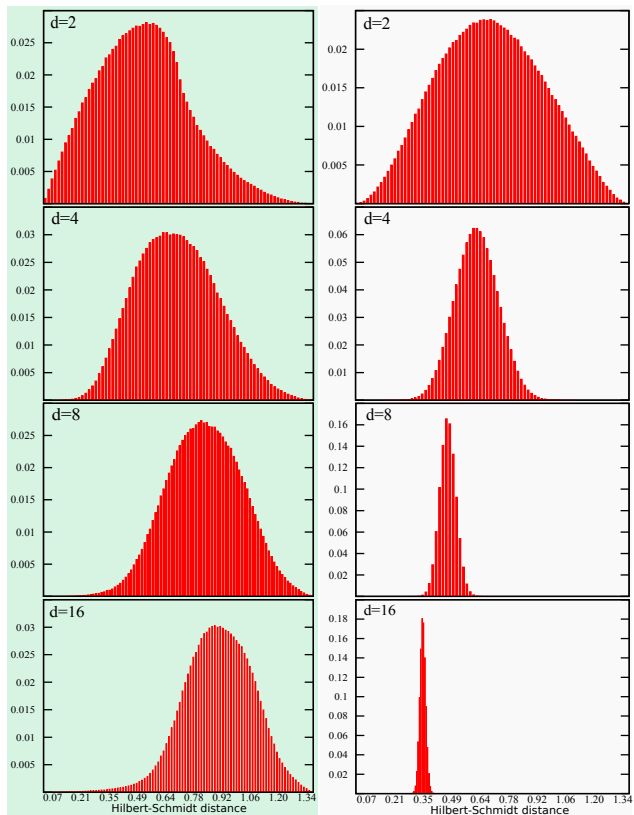}
\par\end{centering}
\protect\caption{(color online) Probability distribution for the different possible
values of the Hilbert-Schmidt distance for one million pairs of quantum
states generated using the standard method (green panel on the left) or
generated via the overparametrized method with $A_{jk}^{r}\mbox{, }A_{jk}^{i}\in[-1,1]$
(gray panel on the right). We see that, in contrast to what happens in the
standard method, there is a too fast concentration of measure in the
OPM as the system's dimension $d$ increases. We observe that although
only some values of $d$ are shown in this figure (1, 2, 3, and 4
qubits), the mentioned effect is smooth and gradual.}
\label{fig2}
\end{figure}

It is known for some time now that in high dimensional spaces random
variables tend to concentrate around their mean values \cite{Ledoux_CoM}.
In the last few years, this phenomenon of concentration of measure,
that is formalized in Levy's Lemma, has gained great importance and
utility in quantum information science (see for instance Ref. \cite{Hayden_CoM}
and the references therein). 

Notwithstanding, as shown in the gray panel on the right hand side of Fig.
\ref{fig2}, the OPM leads to a too fast concentration of measure
for the Hilbert-Schmidt distance in the quantum state space as the
system's dimension $d$ increases. We note that such a concentration
of measure is much more slow in our benchmarking method: the standard
method with the Hurwitz's parametrization for unitary matrices. It
is worth observing that, in part, the shift in the probability distribution
for the HSD observed with the standard method (green panel on the left hand side of
Fig. \ref{fig2}) can be understood as being due to the fact that
as $d$ increases the same number of points will be spread in a ``bigger''
space, diminishing thus the chance for closer pairs of configurations
to be generated.

It is important mentioning that if instead of generating $A$ as described
above, we draw it from the \emph{Ginibre ensemble} \cite{zyczkowski_rrho,zyczkowski_book},
i.e., if we produce $A_{jk}^{r}$ and $A_{jk}^{i}$ using random numbers
normally distributed (and with average equal to zero and variance equal to one),
very similar results are obtained, as is shown in Table \ref{table}.
Thus the effect seems to be a characteristic trait of the overparametrized
method, being independent on how it is applied. 

\begin{table}[h]
\begin{centering}
\begin{tabular}{|c|c|c|c|c|c|c|}
\cline{2-7} 
\multicolumn{1}{c|}{} & \multicolumn{2}{c|}{Uniform} & \multicolumn{2}{c|}{Normal} & \multicolumn{2}{c|}{Standard}\tabularnewline
\hline 
\hline 
$d$ & $\langle d_{hs}\rangle$ & $\Delta d_{hs}$ & $\langle d_{hs}\rangle$ & $\Delta d_{hs}$ & $\langle d_{hs}\rangle$ & $\Delta d_{hs}$\tabularnewline
\hline 
$2$ & $0.697$ & $0.267$ & $0.728$ & $0.267$ & $0.524$ & $0.243$\tabularnewline
\hline 
$4$ & $0.626$ & $0.111$ & $0.655$ & $0.113$ & $0.702$ & $0.213$\tabularnewline
\hline 
$6$ & $0.538$ & $0.063$ & $0.558$ & $0.065$ & $0.794$ & $0.204$\tabularnewline
\hline 
$8$ & $0.476$ & $0.042$ & $0.490$ & $0.043$ & $0.844$ & $0.195$\tabularnewline
\hline 
$10$ & $0.431$ & $0.034$ & $0.442$ & $0.031$ & $0.874$ & $0.190$\tabularnewline
\hline 
$12$ & $0.396$ & $0.023$ & $0.405$ & $0.024$ & $0.894$ & $0.185$\tabularnewline
\hline 
$14$ & $0.369$ & $0.019$ & $0.376$ & $0.019$ & $0.908$ & $0.182$\tabularnewline
\hline 
$16$ & $0.346$ & $0.015$ & $0.352$ & $0.016$ & $0.918$ & $0.179$\tabularnewline
\hline 
\end{tabular}
\end{centering}
\caption{Mean value ($\langle d_{hs}\rangle$) and standard deviation ($\Delta d_{hs}$)
of the Hilbert-Schmidt distance for one million pairs of $d-$dimensional
quantum states randomly generated using the overparametrized method
with uniformly or normally distributed random numbers or generated
using the standard method. }
\label{table}
\end{table}

We also see in Table \ref{table} that, even though the concentration
of measure is ubiquitous, while the width of the probability distribution
for the HSD obtained via the OPM applied to four qubits is less than
$6\mbox{ \%}$ of that obtained in the one-qubit case, for the standard
method the corresponding percentage is almost $74\mbox{ \%}$. We notice
another bold difference between the two methods: as $d$ increases, they shift $\langle d_{hs}\rangle$,
the ``typical'' value of the HSD, in opposite directions (see also Fig. \ref{fig2}).

\subsection{Ginibre and Bures methods}
\label{bures}

\begin{figure}
\begin{centering}
\includegraphics[scale=0.35]{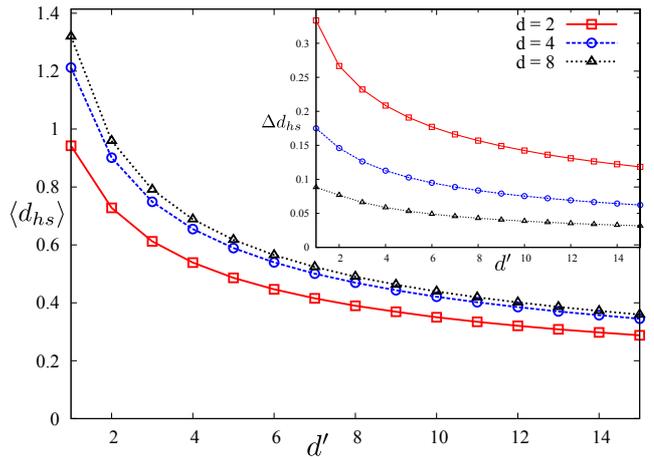}
\par\end{centering}
\protect\caption{(color online) Average of the Hilbert-Schmidt distance, $\langle d_{hs} \rangle$, and the associated standard deviation, $\Delta d_{hs}$, as a function of the Ginibre matrix left dimension $d'$ for some values of its right dimension $d$. A sample with one million pairs of $d\mathrm{x}d$ density matrices was created using the Ginibre method for each pair $(d',d)$. We see that both $\langle d_{hs} \rangle$ and $\Delta d_{hs}$ decrease with $d'$ for a specified value of $d$.}
\label{hsd_induced}
\end{figure}

For completeness, in this sub-section we briefly describe two other methods for RQS generation whose starting point is also the sampling of matrices from the Ginibre ensemble. Let us begin with a generalization of the OPM, that will be named here as the \emph{Ginibre method}. If the $d'\mathrm{x}d$ Ginibre matrix $A$ is a square matrix as the ones considered in Sec. \ref{concentration}, i.e., if $d'=d$, the RQS are generated with a Hilbert-Schmidt measure. On the other hand, in the general case where the number of lines and columns of $A$ need not to coincide, the RQS are said to be generated with an induced measure \cite{zyczkowski_rrho}. Using the Ginibre method to generate a sample with one million pairs of states for each pair $(d',d)$, we show in Fig. \ref{hsd_induced} the dependence with $d'$ of the average and standard deviation of the Hilbert-Schmidt distance for some values of $d$. We see a strong dependence of both quantities with $d'$. This raises an additional practical question about this method. Which value of the Ginibre matrix left dimension $d'$ should be used and how to justify the choice?

\begin{figure}
\begin{centering}
\includegraphics[scale=0.42]{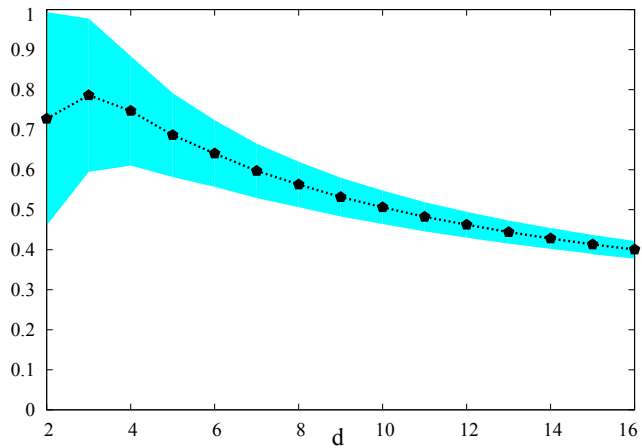}
\par\end{centering}
\protect\caption{(color online) Average Hilbert-Schmidt distance $\langle d_{hs} \rangle$ (black points) as a function of the system dimension $d$. A sample with one million pairs of density matrices was generated, for each value of $d$, using the Bures' method. In the shadowed cyan area are shown values of the HSD standing
between $\langle d_{hs} \rangle - \Delta d_{hs}$ and $\langle d_{hs} \rangle + \Delta d_{hs}$.}
\label{hsd_bures}
\end{figure}

Now we describe the other method, which shall be dubbed as the \emph{Bures' method} because it leads to RQS with a Bures measure. This is accomplished by defining \cite{zyczkowski_bures}
\begin{equation}
\rho=\frac{(\mathbb{I}_{d}+U)AA^{\dagger}(\mathbb{I}_{d}+U^{\dagger})}{\mathrm{Tr}((\mathbb{I}_{d}+U)AA^{\dagger}(\mathbb{I}_{d}+U^{\dagger}))},
\end{equation}
with $A$ being a $d\mathrm{x}d$ Ginibre matrix and $U$ is $d\mathrm{x}d$ a random unitary matrix. It is note worthy that $3d^{2}-d$ real parameters are necessary to create a RQS via this method. For one million pairs of states generated in this way, we show in Fig. \ref{hsd_bures} the center and width of the probability distribution for the Hilbert-Schmidt distance as a function of the system dimension $d$. A behavior similar to that observed for the overparametrized and Ginibre methods, discussed respectively in the last sub-section and in the last paragraph, is seem here. However the rate of concentration of measure is a little less pronounced when compared with that for the OPM. For the Bures' method the width of the probability distribution for four qubits is approximately $8\mbox{ \%}$ of that for one qubit.

\section{Final remarks}
\label{conclusions}

In this article we presented a brief survey of some methods that may be used for the numerical generation of random quantum states. We gave particular emphasis to the overparametrized method, which is frequently used in quantum information science.
After utilizing a qubit system to identify and solve a physically relevant problem related
to the domains of the matrix elements used so far in the literature
in implementations of the OPM, we considered its possible application
for random sampling in high-dimensional quantum systems. In this last
scenario we showed that the overparametrized and related methods lead to a too rapid concentration of
measure that may prevent any fair random sampling of quantum states, even for
quantum systems with moderate dimension. 

\begin{acknowledgements}
This work was supported by the Brazilian funding agencies: Conselho
Nacional de Desenvolvimento Cient\'ifico e Tecnol\'ogico (CNPq), under processes 441875/2014-9 and 303496/2014-2, Instituto Nacional de Ci\^encia e Tecnologia de Informa\c{c}\~ao Qu\^antica (INCT-IQ), under process 2008/57856-6, and Coordena\c{c}\~ao de Desenvolvimento de Pessoal de N\'{i}vel Superior (CAPES), under process 6531/2014-08. I gratefully acknowledge the hospitality of the Laser Spectroscopy Group at the Universidad de la Rep\'{u}blica, Uruguay, where this article was completed. I thank Karol \.{Z}yczkowski for making me aware of Ref. \cite{zyczkowski_asymp}, where (different) asymptotic average values were obtained for different distance measures.
\end{acknowledgements}


\end{document}